\documentclass[aps,pra,twocolumn,amsmath,amssymb,footinbib,showpacs,superscriptaddress]{revtex4-1}

\newcommand{\Jnature}{Nature (London)}
\newcommand{\Jnatphys}{Nature Phys.}

\newcommand{\Jscience}{Science}

\newcommand{\Jprl}{Phys. Rev. Lett.}
\newcommand{\Jpr}{Phys. Rev.}
\newcommand{\Jpra}{Phys. Rev. A}
\newcommand{\Jprb}{Phys. Rev. B}

\newcommand{\JplA}{Phys. Lett. A}

\newcommand{\Jepl}{Europhys. Lett.}
\newcommand{\Jnjp}{New J. Phys.}

\newcommand{\Jepjd}{Eur. Phys. J. D}

\newcommand{\JRepProgPhys}{Rep. Prog. Phys.}

\newcommand{\JjphysA}{J. Phys. A: Math. Theor.}

\newcommand{\Jphystoday}{Phys. Today}

\newcommand{\JZphysB}{Z. Phys. B}

\usepackage[english]{babel}
\usepackage{latexsym}
\usepackage{graphics}
\usepackage{subfigure}
\usepackage{epsfig}
\usepackage{color}
\usepackage{hyperref}
\hypersetup{
colorlinks=true,
citecolor=blue,
linkcolor=red,
urlcolor=black
}

\newcommand{\epsfboxmod}[1]{\epsfbox{#1}}

\newcommand{\infigbis}[2]{          \mbox{ \epsfxsize #1 \epsfboxmod{#2}}
                                      \vspace{-0.8cm}}

\newcommand{\ie}{{i.e.}}
\newcommand{\eg}{{e.g.}}

\renewcommand{\Re}{\mathcal{R}\textrm{\small e}}
\renewcommand{\Im}{\mathcal{I}\textrm{\small m}}

\newcommand{\vect}[1]{\mathbf{#1}}
\newcommand{\av}[1]{\overline{#1}}

\newcommand{\Vr}{V_\textrm{\tiny R}}

\newcommand{\anifact}{\xi}
\newcommand{\sigmar}{\sigma_\textrm{\tiny R}}

\newcommand{\Cor}{C}

\newcommand{\TFCor}{\tilde{C}}

\newcommand{\selfE}{\Sigma}

\newcommand{\Diff}{\vect{D}}
\newcommand{\DiffTensB}{\vect{D}_{\textrm{\tiny B}}}

\newcommand{\Emob}{E_{\textrm{c}}}
\newcommand{\Ebe}{E_{\textrm{be}}}

\newcommand{\ud}{\mathrm{d}}
\newcommand{\vecr}{\textbf{r}}
\newcommand{\veck}{\textbf{k}}

\newcommand{\vecq}{\textbf{q}}

\newcommand{\vecj}{\textbf{j}}

\newcommand{\be}{\begin{equation}}
\newcommand{\ba}{\begin{align}}
\newcommand{\ee}{\end{equation}}
\newcommand{\ea}{\end{align}}

\newcommand{\Gr}{G}
\newcommand{\Ga}{G^{\dagger}}

\newcommand{\epsfree}[1]{\epsilon_0(#1)}
\newcommand{\grad}{\boldsymbol{\nabla}}
\begin{document}

\title{Anderson Localization of Matter Waves in 3D Anisotropic Disordered Potentials}

\author{Marie Piraud}
\affiliation{Laboratoire Charles Fabry, Institut d'Optique, CNRS, Univ Paris Sud
11, 2 avenue Augustin Fresnel, F-91127 Palaiseau cedex, France}
\affiliation{Universit\'e Grenoble 1/CNRS, LPMMC UMR 5493, B.P.~166, F-38042 Grenoble,
France}
\affiliation{Department of Physics and Arnold Sommerfeld Center for Theoretical Physics, Ludwig-Maximilians-Universit\"at M\"unchen, D-80333 M\"unchen, Germany}

\author{Laurent Sanchez-Palencia}
\affiliation{Laboratoire Charles Fabry, Institut d'Optique, CNRS, Univ Paris Sud
11, 2 avenue Augustin Fresnel, F-91127 Palaiseau cedex, France}

\author{Bart van Tiggelen}
\affiliation{Universit\'e Grenoble 1/CNRS, LPMMC UMR 5493, B.P.~166, F-38042 Grenoble,
France}

\date{\today}

\begin{abstract}
Using a cut-off free formulation of the coherent transport theory, we show that the interference terms at the origin of localization strongly affect the transport anisotropy.
In contrast to the common hypothesis, we then find that the anisotropies of incoherent and coherent diffusion
are significantly different, in particular at criticality.
There, we show that the coherent transport anisotropy is mainly determined
by the properties of the disorder-averaged effective scattering medium
while the incoherent transport contributions become irrelevant.

\end{abstract}


\maketitle

 \section{Introduction}

The propagation of a coherent wave in a disordered medium is strongly affected by interference
of the various multiple-scattering paths, which may suppress or even cancel diffusion.
This phenomenon, known as Anderson localization (AL), is a widely studied problem
at the frontier of condensed-matter and wave physics.
Anderson localization has now been observed in a variety of systems
with
electromagnetic waves~\cite{wiersma1997,chabanov2000,storzer2006},
acoustic waves~\cite{hu2008},
and ultracold matterwaves~\cite{billy2008,roati2008,chabe2008,kondov2011,jendrzejewski2012,mcgehee2013,semeghini2014}.
Basic knowledge of AL relies on the discrete Anderson model~\cite{anderson1958,mackinnon1983,slevin2001}
and the one-parameter scaling theory~\cite{abrahams1979}, which in particular predicts a universal localization transition in dimension $d>2$.
Recent progress on control of disordered systems~\cite{lagendijk2009,aspect2009,lsp2010,modugno2010,shapiro2012} triggered a renewed interest for a refined, microscopic understanding of AL,
which cannot be described in the framework of the universal scaling theory.

An example where a microscopic theory is vital is that of localization in anisotropic media, which are relevant to 
  MOSFETs~\cite{bishop1984},
  liquid crystals~\cite{kao1996,wiersma1999},
  phosphides~\cite{johnson2002},
  or ultracold atoms~\cite{kondov2011,jendrzejewski2012,mcgehee2013,piraud2012a,piraud2013b,plodzien2011,piraud2012b,semeghini2014},
for instance.
These systems may be cast in two classes of anisotropy,
either resulting from isotropic scatterers embedded in an anisotropic underlying medium (mass anisotropy)~\cite{kaas2008} or resulting from anisotropic disorder in an isotropic medium (disorder anisotropy).
For these systems, basic knowledge presently relies on the extension of the on-shell self-consistent theory to anisotropic media~\cite{woelfle1984}. The latter predicts that the interference terms have the same anisotropy as the incoherent propagation terms, so that the anisotropies of localization and classical diffusion are equal.
However, this prediction relies on the introduction of an
elliptic cut-off whose anisotropy is arbitrarily chosen as the inverse of the incoherent transport mean free path.

In this paper, using a cut-off free formulation of the coherent transport theory, we show that interference terms at the origin of AL strongly affect the transport anisotropy.
The solution of the transport equations shows that the anisotropies of incoherent and coherent diffusion
are significantly different, especially near the mobility edge (ME).
This invalidates the use of the elliptic cut-off used so far.
Our work indicates that the anisotropy in coherent diffusion is mainly determined by 
the properties of the average effective medium.
Hence, contrary to the usual hypothesis, the interference terms significantly compensate for
the large anisotropy predicted by incoherent Boltzmann transport theory.
Our results provide insight into the theory of AL in anisotropic disorder and should significantly affect the interpretation of experimental data.

 \section{Quantum transport}
\subsection{Formalism}
The coherent transport equations for a wave in an anisotropic disordered medium are obtained from the exact Bethe-Salpeter equation~\cite{rammer1998} by generalizing the approach of Ref.~\cite{yedjour2010} to the anisotropic case.
Briefly speaking (see Appendix~\ref{app:form} for details), the density propagation of a matterwave of energy $E$ and wavevector $\veck$ in a
disordered medium is described in the long time and large distance limit by the static current vertex function $\vecj(E,\veck)$, which obeys the closed equation
\be \label{eq:j-app}
\vecj(E,\veck)=\vecj_0(E,\veck) + \vert \av{\Gr}(E,\veck) \vert^2 \int \frac{\ud \veck'}{(2\pi)^d} \, \vecj(E,\veck') \, U_{\veck,\veck'}(E)
\ee
where
$\vecj_0(E,\veck)=\left[\veck+\grad_{\veck} \Re\selfE(E,\veck) \right] \left[\Im\av{\Gr}(E,\veck)\right]^2 
-\frac{1}{2} \Re\av{\Gr}(E,\veck) \Im\av{\Gr}(E,\veck) \grad_{\veck} \Im\selfE(E,\veck)$,
$\av{\Gr}(E,\veck)$ is the disorder average of the single-particle Green function,
$\selfE(E,\veck)$ is the associated self-energy,
and $U_{\veck,\veck'}(E)= U_{\veck,\veck'}(E,\omega=0,\vecq=\mathbf{0})$ is the static irreducible vertex function ($\omega$ and $\vecq$ are the conjugate variables of time and space, respectively).
The components of the diffusion tensor $\Diff(E)$ are then given by the Kubo formula
\be \label{eq:kubo}
D^{uv}(E)=\frac{\hbar}{m}\frac{1}{\pi N(E)} \int \frac{\ud \veck}{(2\pi)^d} \, k_u j_v(E,\veck),
\ee
with $N(E)=\int\frac{\ud \veck}{(2\pi)^d}\, \frac{-\Im\av{\Gr}(E,\veck)}{\pi}$ the density of states per unit volume.

\subsection{Interference term}
In the weak localization regime, we retain the following form for the irreducible vertex function
\be \label{eq:UTot}
U_{\veck,\veck'}(E)=U^{\textrm{\tiny B}}_{\veck,\veck'}+U^{\textrm{\tiny MC}}_{\veck,\veck'}(E),
\ee
where $U^{\textrm{\tiny B}}_{\veck,\veck'}=\TFCor(\veck-\veck')$ is the vertex associated with incoherent (Boltzmann) transport, with $\TFCor(\veck)$ the power spectrum 
(fluctuations in the potential energy)~\footnote{Here, we use $\TFCor(\vecq,\omega) \equiv \int \ud\vecr \ud t\ \Cor(\vecr,t) \exp[-i (\vecq\cdot\vecr - \omega t)]$},
and $U^{\textrm{\tiny MC}}_{\veck,\veck'}$
is the leading interference contribution expressed by the maximally-crossed diagrams~\cite{yedjour2010}, 
\be \label{eq:UMaxCross}
U^{\textrm{\tiny MC}}_{\veck,\veck'}(E)=\frac{2}{\pi N(E)}\frac{\left[ \Im\selfE(E,\frac{\veck-\veck'}{2}) \right]^2}{(\veck+\veck') \cdot \Diff(E) \cdot (\veck+\veck')}.
\ee
Note that the diffusion tensor $\Diff(E)$ is included self-consistently in Eqs.~(\ref{eq:j-app}) and (\ref{eq:kubo}),
an approximation which gives quantitative estimates in good agreement with numerical calculations~\cite{kroha1990}.
In standard (on-shell) self-consistent theory, the pole of Eq.~\eqref{eq:UMaxCross} necessitates the introduction of an ultraviolet cut-off in the integral in the right-hand side of Eq.~\eqref{eq:j-app}. Here, by keeping the full $\veck$ dependence of the vertex function $\vecj(E,\veck)$, this artificial divergence does not appear~\cite{yedjour2010}.

\subsection{Disorder correlation function}
For simplicity, we assume that the disorder $V(\vecr)$ is Gaussian-distributed.
Its statistical properties are fully characterized by the two-point correlation function $\Cor(\vecr)=\av{V(\mathbf{0}) V(\vecr)}$,
where we have set the zero of energies such that $\av{V}=0$.
In order to
carry out a number of integrations analytically,
it is convenient to use a correlation function with uniaxial symmetry
around, say, axis $z$,
and to adopt
$\Cor(\vecr)=\Vr^2 \exp\{-(x^2+y^2+z^2/\anifact^2)/\sigmar^2\}$,
where
$\Vr$ is the amplitude of the disorder, $\sigmar$ its typical correlation length, and $\anifact$ its anisotropy.
It is the simplest model of anisotropic disorder and it is relevant to ultracold-atom experiments,
see \eg\ Refs.~\cite{kondov2011,mcgehee2013} where $\anifact$ varies from 6 to 20.
When $\anifact>1$, this correlation function corresponds to a correlated disorder made of 'grains', of typical transverse size $\sigmar$ and elongated in the $z$ direction.

\begin{figure*}[t!]
\begin{center}
\infigbis{54em}{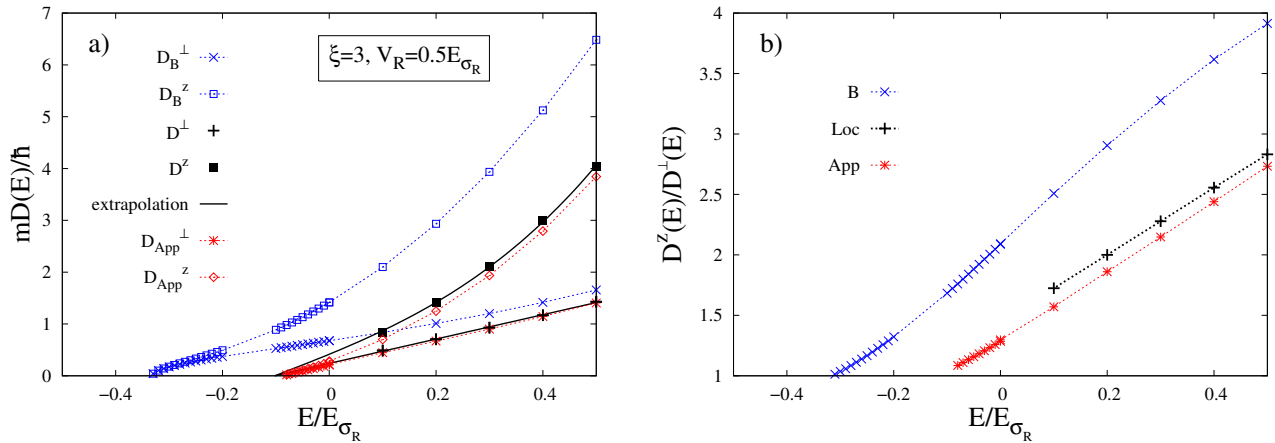}
\end{center}
\vspace{0.1cm}
\caption{
\label{fig:DiffCoeffs}
(Color online)
Quantum coherent diffusion versus Boltzmann diffusion.
(a)~Diffusion coefficients in the transverse plane $\perp=\{x,y\}$ and in the longitudinal direction $z$ computed from the numerical solution of Eq.~(\ref{eq:j-app}) for our model of disorder with $\anifact=3$ and $\Vr=0.5 E_{\sigma}$.
Shown are the results for
the incoherent Boltzmann approximation ($U\simeq U^{\textrm{\tiny B}}$, blue points),
the quantum diffusion ($U=U^{\textrm{\tiny B}}+U^{\textrm{\tiny MC}}$, black points),
and the approximation underlying Eq.~(\ref{eq:approx}) [red points].
The black solid line shows the extrapolation $D^{\perp}(E) \propto (E-\Emob)$ and $D^z(E) \propto (E-\Emob)[1+A^{z}(E-\Emob)^2]$, which gives $\Emob=-0.102 E_{\sigma}$.
(b)~Anisotropy of the quantum and Boltzmann diffusion tensors, $D^z(E)/D^{\perp}(E)$.
}
\end{figure*}

 \section{Localization Regime}

 \subsection{Exact numerical computation}
The first step in the calculation consists in determining the average Green function, which enters into Eq.~(\ref{eq:j-app}).
It reads $\av{\Gr}(E,\veck)=[E-\hbar^2 k^2/2m-\selfE(E,\veck)]^{-1}$,
where $\selfE$ is given by the solution of the Dyson equation.
In order to study the fundamental effects due to disorder anisotropy we compute 
$\selfE$ in the self-consistent Born approximation (SCBA), which reads
\be \label{eq:scba}
\selfE(E,\veck)=\int \frac{\ud \veck'}{(2\pi)^d}\, \frac{\TFCor(\veck-\veck')}{E-\hbar^2 k'^2/2m-\selfE(E,\veck')}.
\ee
This approximation reproduces qualitatively single-particle features such as the density of states for Gaussian disorder.
It is therefore well suited for understanding basic effects of geometrical disorder anisotropy
as long as the disorder is not too large~\cite{delande2014}.
The major benefit of the SCBA method compared to the on-shell first-order perturbation theory is that it 
properly accounts for the momentum decay of the self-energy $\selfE(E,\veck)$.

We now numerically solve Eq.~(\ref{eq:j-app}) for the current vertex function $\vecj$ by iteration (see Appendix~\ref{app:num-tech} for details)
 and insert the result into Eq.~(\ref{eq:kubo}) to determine the diffusion tensor.
Due to parity and uniaxial symmetries,
the diffusion tensor admits $x$, $y$, and $z$ as eigendirections and
is isotropic in the $(x,y)$ plane.
We call $D^{\perp}$ the corresponding diffusion coefficient
 and $D^z$ the coefficient along the $z$ direction.
In Fig.~\ref{fig:DiffCoeffs}(a) we show the results for the diffusion coefficients (black points) for
$\anifact=3$ and $\Vr=0.5 E_{\sigma}$ with $E_{\sigma}=\hbar^2/m\sigmar^2$
and in Fig.~\ref{fig:DiffCoeffs}(b) the anisotropy $D^z/D^{\perp}$.
For comparison, the corresponding values obtained in the Boltzmann approximation,
\ie\ if only $U^{\textrm{\tiny B}}$ is retained in Eq.~(\ref{eq:UTot}),
are also shown (blue points).
We find that, as expected, both diffusion tensors are anisotropic with larger values in the $z$ direction, and that both decrease with decreasing energy.
The Boltzmann diffusion tensor $\DiffTensB(E)$ "trivially" vanishes at the point where the density of states
vanishes (hereafter called band edge), 
$\Ebe \simeq -0.35 E_{\sigma}$.
As expected in the framework of AL, the quantum diffusion tensor $\Diff(E)$
vanishes well before $\DiffTensB(E)$.
The energy dependence of $\Diff(E)$ (black points) points toward a vanishing point at $\Emob > \Ebe$,
for both $D^{\perp}$ and $D^z$.
It corresponds to the
ME, the states with energy $\Ebe<E<\Emob$ being localized.
The iterative calculation suffers from the usual critical slowing down phenomenon, which prevents us from performing exact calculations very close to the ME.
In order to find the ME, we use
the extrapolating functions $D^z(E) \propto (E-\Emob)[1+A^{z}(E-\Emob)^2]$ and $D^{\perp}(E) \propto (E-\Emob)$, which fit the data properly.
This extrapolation is represented by the solid black line in Fig.~\ref{fig:DiffCoeffs}(a), and gives $\Emob \simeq -0.1 E_{\sigma}$.

 \subsection{Approximation}
In order to avoid the critical slowing down in solving Eq.~(\ref{eq:j-app}),
it is worth simplifying the problem.
It is vital to avoid any on-shell description
since they would introduce artificial divergencies at short length scales.
To do so, we assume here that the $\veck$-dependence of $\vecj(E,\veck)$ is equal to the one of  $\vecj_{\textrm{\tiny 0}}(E,\veck)$, which amounts to assuming that the momentum distribution of the atoms at fixed energy does not change when interference terms are included.
We have checked that this is indeed a fairly good approximation, as for isotropic models of disorder~\cite{yedjour2010}.
Then, assuming $\vecj(E,\veck) = \Diff(E) \Diff_{\textrm{\tiny 0}}^{-1}(E) \, \vecj_{\textrm{\tiny 0}}(E,\veck)$ and $\vecj_{\textrm{\tiny B}}(E,\veck) = \DiffTensB(E) \Diff_{\textrm{\tiny 0}}^{-1}(E) \, \vecj_{\textrm{\tiny 0}}(E,\veck)$, we get the following closed equation for the diffusion tensor
\begin{align} \label{eq:approx}
D^{u}(E) &\simeq D^{u}_{\textrm{\tiny B}}(E) \Bigg\{ 1 +
\frac{2\hbar^2 D^u(E)}{m^2 \pi^2 N(E)^2 D^u_{\textrm{\tiny 0}}(E)^2} \\
&\times \int \frac{\ud \veck}{(2\pi)^d} \, \frac{\ud \veck'}{(2\pi)^d} \, k_u \vert \av{\Gr}(E,\veck) \vert^2 \, j_{u,\textrm{\tiny 0}}(E,\veck') \nonumber \\
& \frac{\left[ \Im\selfE(E,\frac{\veck-\veck'}{2}) \right]^2}{(\veck+\veck') \cdot \Diff(E) \cdot (\veck+\veck')} \Bigg\}, \nonumber
\end{align}
where $u \in \{x,y,z\}$ are the three directions.
Notice that there is a great gain in computational time as we have simplified an iterative problem for a vector of three-dimensional functions $\vecj(E,\veck)$ (respectively two-dimensional in the uniaxial case we consider here) to an iterative problem for a tensor with only three (respectively two) parameters.
This allows us to compute the diffusion tensor for energies down to the mobility edge $\Emob$.

The results of this approximation are shown in Fig.~\ref{fig:DiffCoeffs}.
We find that the eigenvalues of the full diffusion tensor [Fig.~\ref{fig:DiffCoeffs}(a)] and the anisotropy [Fig.~\ref{fig:DiffCoeffs}(b)] are fairly well reproduced within the approximation.
In particular, we recover that the eigenvalues of the diffusion tensor all vanish at the same critical point.
This may be understood by a dimensionality reduction argument.
Should localization occur in one or two directions only, the remaining (transverse) dynamics would be reduced to a dimension lower than three.
It would then systematically lead to localization also in the transverse directions.
The same conclusion can also be found from Eq.~(\ref{eq:approx}).
Assume that the diffusion coefficient vanishes in the eigendirection $u$, but not in the others.
Then the integral in Eq.~(\ref{eq:approx}) would be finite, so that $D^{u}(E) = D^{u}_{\textrm{\tiny B}}(E)$.
This would be clearly inconsistent since $D^{u}(E)=0$ by hypothesis while $D^{u}_{\textrm{\tiny B}}(E) \neq 0$.
Using the approximation, we locate the ME at $\Emob =-0.09 E_{\sigma}$ for $\Vr=0.5E_{\sigma}$ and $\anifact=3$,
which is in good agreement with the value we found by extrapolating the exact numerical results, down to $10\%$.

\begin{figure*}[t!]
\begin{center}
\infigbis{54em}{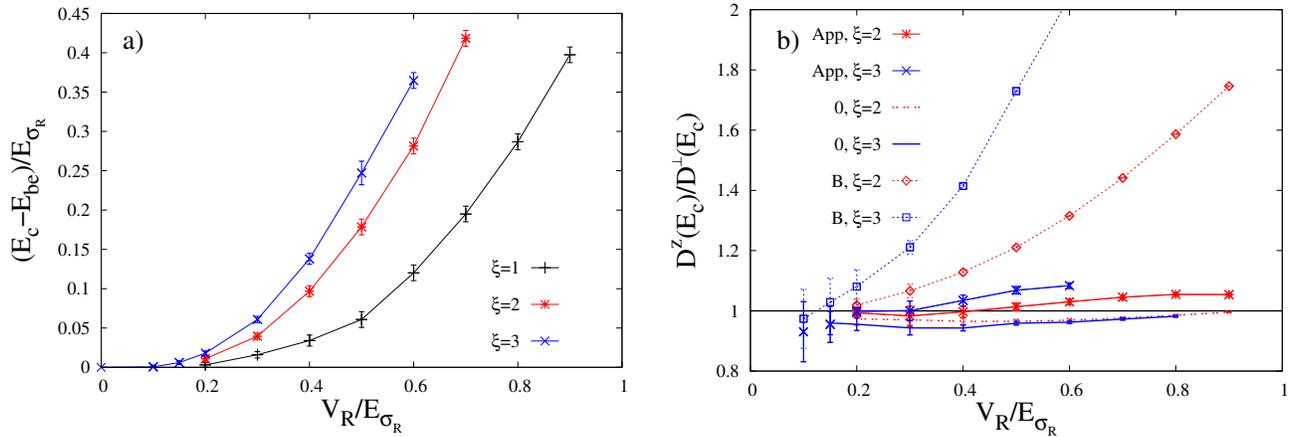}
\end{center}
\vspace{0.1cm}
\caption{
\label{fig:Emob}
(Color online) (a) Position of the mobility edge with respect to the band edge $\Emob-\Ebe$ as a function of the amplitude of the disorder in approximation~(\ref{eq:approx}), for geometrical anisotropies $\anifact=1$, 2 and 3 (black, red and blue data points respectively).
(b) Anisotropy of the diffusion tensor at the mobility edge found within the approximation for $\anifact=2$ and 3 (red and blue data points respectively), compared with the anisotropy of $\Diff_0$ and $\DiffTensB$ at the same energy (see legend).
}
\end{figure*}

 \subsection{Mobility edge}
The ME, measured with respect to the band edge, $\Emob-\Ebe$, is shown in Fig.~\ref{fig:Emob}(a)
versus the disorder amplitude $\Vr$, for various geometrical anisotropies.
For low values of $\Vr$, the ME is very close to, but above, the band edge, well below the average value of the disorder, $\av{V}=0$. When $\Vr$ increases, the ME moves away from the band edge.
Moreover,
increasing the disorder anisotropy increases the ME at fixed $\Vr$
and therefore tends to favor localization.
This is intuitive since
a strongly elongated disorder ($\anifact \gg 1$)
tends to reduce the relative importance of the $z$ direction.
We also define the parameter
$kl(E) \equiv m \textrm{Tr} \left[ \Diff_0(E) \right]/\hbar$,
which coincides with the Ioffe-Regel parameter in the case of isotropic short-range disorder~\cite{ioffe1960,*mott1979},
since $l(E)$ is then the scattering mean-free path.
When computed at the ME, we find that $kl$ is of order unity, it weakly depends on the disorder anisotropy,
and it shows a slight increase with the disorder amplitude similar to that found for isotropic disorder~\cite{yedjour2010}.
It thus provides a generalization of the Ioffe-Regel criterion to the anisotropic case.

 \subsection{Transport anisotropies}
We now discuss the transport anisotropies, $D^z_{\textrm{\tiny B}}/D_{\textrm{\tiny B}}^{\perp}$ and $D^z/D^{\perp}$, which are shown in Fig.~\ref{fig:DiffCoeffs}(b) versus the matterwave energy $E$.
We find that both decrease with energy,
similarly as found in the on-shell approximation~\cite{piraud2012a,*piraud2013b}.
The Boltzmann diffusion tensor becomes isotropic only at very low energy when it approaches the band edge.
This is consistent with the fact that our model has a nondiverging infrared limit,
$\lim_{\veck \rightarrow 0} \TFCor(\veck) = \textrm{const}$.
Then, for low energy, all integrals are dominated by small wavevectors $\veck \lesssim \sigmar^{-1}$,
which are almost insensitive to the details of the correlation function, in particular to its anisotropy.
Most important, we find that the anisotropy of the quantum diffusion tensor $\Diff(E)$ is significantly smaller than that of the Boltzmann diffusion tensor $\DiffTensB(E)$.
It shows that the localization corrections strongly affect the transport anisotropy,
leading to a significant reduction for our model of disorder.
For instance in Fig.~\ref{fig:DiffCoeffs}(b), when $E\rightarrow\Emob$,
the coherent transport anisotropy $D^z/D^{\perp}$ strongly decreases and becomes of the order of 1,
while the anisotropy of the incoherent transport $D^{z}_{\textrm{\tiny B}}/D^{\perp}_{\textrm{\tiny B}}$
is of the order of 1.75.
This effect is the main outcome of our approach, which is free of any cut-off of arbitrary anisotropy, unlike previous work~\cite{woelfle1984,piraud2012a,*piraud2013b}.

Figure~\ref{fig:Emob}(b) shows the anisotropies of the diffusion tensors at the ME,
where the deviations are more pronounced.
Isotropic diffusion tensors are found only in the white-noise limit $\Vr \ll E_{\sigmar}$~\footnote{The limit $\Vr/E_{\sigmar} \rightarrow 0$ is obtained by taking $\sigmar \rightarrow 0$ while $\Vr^2 \sigmar^3$ is fixed.
The correlation function can then be approximated by
$C(\vecr) \simeq \Vr^2 \sigmar^3 \delta(\vecr/\sigmar)$.}.
Conversely, for larger values of $\Vr$, finite $\veck$ components of the disorder contribute.
Then, we find that the self-energy $\selfE$ becomes only slightly anisotropic and thus,
that the anisotropy of $\Diff_0$ is weakly affected.
In contrast, the anisotropy of $\DiffTensB$ is significantly affected.
This is due to the second term in Eq.~(\ref{eq:j-app}), which couples $\vecj(E,\veck)$ to the disorder power spectrum $\TFCor(\veck)$ in the Boltzmann approximation.
Most importantly, we find that the anisotropy of the quantum diffusion tensor $\Diff$ is significantly smaller than that of $\DiffTensB$. It confirms the strong renormalization of the transport anisotropy by localization corrections.
To understand this behavior, let us rely on Eq.~(\ref{eq:approx}), which relates $\Diff$
to $\DiffTensB$ and properties of the disorder-average effective scattering medium.
Conversely, away from the ME, the localization corrections [second term in the bracket of Eq.~(\ref{eq:approx})] vanish and $\Diff \simeq \DiffTensB$.
When approaching the ME localization corrections increase,
which tend to lower the anisotropy of $\Diff$.
More precisely, at criticality, $\Diff \simeq 0$ while $\DiffTensB \neq 0$.
Hence, $\Diff$ is solely determined by the cancellation of the term inside the brackets in Eq.~(\ref{eq:approx}) and
$\DiffTensB$ becomes irrelevant. We can roughly estimate the anisotropy at criticality by neglecting the width of the pole in Eq.~(\ref{eq:approx}) and estimating all integrals in the range where their integrands are maximal.
This yields
${D^z(\Emob)}/{D^{\perp}(\Emob)} \sim {D^z_0(\Emob)}/{D^{\perp}_0(\Emob)}
$
for short-range correlations,
which is consistent with the weak anisotropy of the quantum diffusion tensor at criticality found numerically in Fig.~\ref{fig:Emob}(c).

 \section{Conclusion}
In summary, we have shown that localization significantly affects transport anisotropy, especially near the ME. Our results show that the anisotropy of the coherent diffusion tensor is mainly determined by the properties of the disorder-averaged effective scattering medium while the incoherent transport contributions are 
significantly compensated by interference when approaching the ME.
More precisely, we find a small but finite anisotropy of transport near the ME.
Those results are obtained in the framework of the approximation leading to Eq.~(\ref{eq:approx}),
which is satisfactory for the parameters used in this work. However, it slightly underestimates the anisotropy.
Relaxing this condition may slightly affect the coherent transport anisotropy.
Our results indicate that a significant deviation from isotropy could be found
in the case of strongly anisotropic scattering properties,
a case that should probably be investigated by going beyond the SCBA for computing properties of the scattering medium.
Our results provide insights into the theory of Anderson localization in anisotropic media
and are directly relevant to experiments on ultracold atoms in three-dimensional anisotropic disorder~\cite{kondov2011,jendrzejewski2012,mcgehee2013,semeghini2014}.
In the future, it would be interesting to compare our predictions to direct numerical calculations using for instance an extension of the method used in Ref.~\cite{delande2014} to anisotropic disorder.

 \section{Acknowledgments}
We thank Serguey Skipetrov for enlightening discussions.
This research was supported by the 
European Research Council (Contract No.\ 256294),
the Agence Nationale de la Recherche (Contract No.\ ANR-08-blan-0016-01),
the Minist\`ere de l'Enseignement Sup\'erieur et de la Recherche,
and
the Institut Francilien de Recherche sur les Atomes Froids (IFRAF).
We acknowledge GMPCS computing facilities of the LUMAT federation and
HPC resources from GENCI-IDRIS (Grants No. 2013057143 and No. 2014057143).

\appendix

\section{Quantum transport formalism \label{app:form}}

The two-particle Green function $\Phi_{\veck,\veck'}(\vecq,\omega, E)$ describes the density propagation of a wave in a disordered medium.
It is defined as
$\Phi_{\veck,\veck'}(\vecq,\omega, E)\equiv 
\av{ \langle \veck_+ \vert \Gr(E_+) \vert \veck'_+ \rangle 
\langle \veck'_- \vert \Ga(E_-) \vert \veck_- \rangle}$
with $\Gr$ the retarded Green operator,
$\veck_\pm \equiv \veck \pm \vecq/2$ and $\veck_\pm' \equiv \veck' \pm \vecq/2$,
$E_\pm \equiv E \pm \hbar\omega/2$,
and $(\vecq$, $\omega)$ the Fourier conjugates of the space and time variables~\footnote{Here, we use $\tilde{f}(\vecq,\omega) \equiv \int \ud\vecr \ud t\ f(\vecr,t)
  \exp[-i (\vecq\cdot\vecr - \omega t)]$.}.

It is governed by the Bethe-Salpeter equation, which can be formally written as~\cite{rammer1998}
\be \label{eq:BSE}
\Phi = \av{\Gr} \otimes \av{\Ga} + \av{\Gr} \otimes \av{\Ga} \, \mathrm{U} \, \Phi \,.
\ee
The first term in Eq.~(\ref{eq:BSE}) describes uncorrelated propagation of the field and its conjugate in the disordered medium. The second term
involves the vertex function $\mathrm{U}$, which includes all irreducible scattering diagrams, and
accounts for all correlations in the density propagation.
The associated quantum kinetic equation reads~\cite{vollhardt1980a,rammer1998}
\begin{align} \label{eq:QKE}
& \left[ \omega - \vecq \cdot \grad_{\veck} \epsfree{\veck} \right] \Phi_{\veck,\veck'}(\vecq,\omega, E) =\\
& \; \; \; \; \; \; \; \; \; \; \; \; \; \; \; \; \; \; \; \; -(2 \pi)^d \delta(\veck-\veck') \Delta \av{\Gr}_{\veck} (E,\omega,\vecq) \nonumber\\
&+ \int \frac{\ud \veck''}{(2\pi)^d} \, U_{\veck'',\veck}(E,\vecq,\omega) \big[ \Delta \av{\Gr}_{\veck''} (E,\omega,\vecq) \, \Phi_{\veck,\veck'}(\vecq,\omega, E) \nonumber\\
& \; \; \; \; \; \; \; \; \; \; \; \; \; \; \; \; \; \; \; \; -\Delta \av{\Gr}_{\veck} (E,\omega,\vecq) \, \Phi_{\veck'',\veck'}(\vecq,\omega, E) \big] \,, \nonumber
\end{align}
where $\epsfree{\veck}$ is the free dispersion relation ($\epsfree{\veck}=\hbar^2 k^2/2m$ in our case, for particles in free space) and
$\Delta \av{\Gr}_{\veck} (E,\omega,\vecq) =  \av{\Gr}(E_+,\veck_+) - \av{\Ga}(E_-,\veck_-)$.
The conservation of the total quantum probability gives the Ward identity which relates the vertex function to the single-particle self-energy $\selfE$ associated to the average Green function $\av{\Gr}$:
\be \label{eq:Ward}
\Delta \selfE_{\veck} (E,\omega,\vecq) = \int \frac{\ud \veck'}{(2\pi)^d} \, U_{\veck,\veck'}(E,\vecq,\omega) \Delta \av{\Gr}_{\veck'} (E,\omega,\vecq).
\ee

The conservation of quantum probability, through the Ward identity, guarantees a diffusion pole.
In the long time ($\omega \rightarrow 0$) and large distance ($\vecq \rightarrow 0$) limit, one can therefore write $\Phi$ as~\cite{barabanenkov1991,rammer1998}:
\be
\Phi_{\veck,\veck'}(E,\omega,\vecq)=\frac{-2}{\int\frac{\ud \veck}{(2\pi)^d}\, \Im\av{\Gr}(E,\veck)} \, \frac{\phi(E,\veck,\vecq)\phi(E,\veck',\vecq)}{-i\omega+\vecq \cdot \Diff(E) \cdot \vecq}
\ee
where $\Diff(E)$ is the diffusion tensor, and $\phi(E,\veck,\vecq)$ is the eigenfunction of the Bethe-Salpeter equation associated with the hydrodynamic diffusion [which corresponds to eigenvalue 0 as $(\vecq, \omega \to 0$)].
The linearization of $\phi$ at small $\vecq$ gives
$\phi(E,\veck,\vecq)=-\Im\Gr(E,\veck)-i\vecq\cdot \vecj(E,\veck)$~\cite{yedjour2010},
where $\vecj$ is the static current vertex function~\cite{woelfle1984,mahan2000}.

Fick's law, which relates the diffusive flux to the concentration gradient in the diffusive regime ($\omega,\vecq \rightarrow 0$) reads $\boldsymbol{\mathcal{J}}(E,\omega,\veck,\vecq)=-i \Diff(E) \vecq \mathcal{P}(E,\omega,\veck,\vecq)$ in Fourier space.
The current density reads $\boldsymbol{\mathcal{J}}(E,\omega,\veck,\vecq)=\int \frac{\ud \veck'}{(2\pi)^d} \, \veck' \Phi_{\veck,\veck'}(E,\omega,\vecq)$, and the probability density $\mathcal{P}(E,\omega,\veck,\vecq)=\int \frac{\ud \veck'}{(2\pi)^d} \, \Phi_{\veck,\veck'}(E,\omega,\vecq)$.
It gives the Kubo formula for the diffusion tensor
\be
D^{uv}(E)=\frac{\hbar}{m}\frac{1}{\pi N(E)} \int \frac{\ud \veck}{(2\pi)^d} k_u j_v(E,\veck).
\label{eq:kubo-app}
\ee

Expanding the quantum kinetic equation~(\ref{eq:QKE}) into terms linear in $\vecq$ gives
\begin{align}
\vecj(E,\veck)=&\frac{\veck}{2} \vert \av{\Gr}(E,\veck) \vert^2 -\frac{1}{2} \grad_{\veck} \Re\av{\Gr}(E,\veck) \\
+&\vert \av{\Gr}(E,\veck) \vert^2 \int \frac{\ud \veck'}{(2\pi)^d} \, \vecj(E,\veck') \, U_{\veck,\veck'}(E,\mathbf{0}) \nonumber\\
+&\frac{1}{2} \vert \av{\Gr}(E,\veck) \vert^2 \int \frac{\ud \veck'}{(2\pi)^d} \, \grad_{\veck} \Re\av{\Gr}(E,\veck') \, U_{\veck,\veck'}(E,\vecq). \nonumber
\end{align}
Developing also the Ward identity~(\ref{eq:Ward}) permits us to find
\begin{align}
\vecj(E,\veck)=&\frac{\veck}{2} \vert \av{\Gr}(E,\veck) \vert^2 -\frac{1}{2} \grad_{\veck} \Re\av{\Gr}(E,\veck)\\
+&\frac{1}{2} \vert \av{\Gr}(E,\veck) \vert^2 \, \grad_{\veck} \Re\selfE(E,\veck) \nonumber\\
+&\vert \av{\Gr}(E,\veck) \vert^2 \int \frac{\ud \veck'}{(2\pi)^d} \, \vecj(E,\veck') \, U_{\veck,\veck'}(E,\mathbf{0}) \nonumber\\
-i \vert \av{\Gr}&(E,\veck) \vert^2 \int \frac{\ud \veck'}{(2\pi)^d} \, \Im\av{\Gr}(E,\veck') \, \grad_{\vecq} U_{\veck,\veck'}(E,\vecq)\vert_{\vecq=0} \ . \nonumber
\end{align}
Finally, we find the following expression:
\begin{align} \label{eq:j-full}
\vecj(E,\veck)&=\vecj_0(E,\veck)\\
&+ \vert \av{\Gr}(E,\veck) \vert^2 \int \frac{\ud \veck'}{(2\pi)^d} \, \vecj(E,\veck') \, U_{\veck,\veck'}(E,\mathbf{0}) \nonumber \\
-i \vert \av{\Gr}&(E,\veck) \vert^2 \int \frac{\ud \veck'}{(2\pi)^d} \, \Im\av{\Gr}(E,\veck') \, \grad_{\vecq} U_{\veck,\veck'}(E,\vecq)\vert_{\vecq=0} \ . \nonumber
\end{align}
with 
$\vecj_0(E,\veck)=\left[\veck+\grad_{\veck} \Re\selfE(E,\veck) \right] \left[\Im\av{\Gr}(E,\veck')\right]^2 
-\frac{1}{2} \Re\av{\Gr}(E,\veck) \Im\av{\Gr}(E,\veck) \grad_{\veck} \Im\selfE(E,\veck) \nonumber$.
In this paper, we have neglected the last term in Eq.~(\ref{eq:j-full}) as $\grad_{\vecq} U_{\veck,\veck'}(E,\vecq)$ vanishes in the Boltzmann approximation, and is less divergent than $U_{\veck,\veck'}(E,\vecq)$ itself for the maximally-crossed diagrams.

\section{Numerical method \label{app:num-tech}}

When solving Eq.~(\ref{eq:j-app}) for the current vertex function $\vecj(E,\veck)$, we proceed by iterations.
At stage $n \geq 1$ we compute $\vecj^{(n)}(E,\veck)$ from $\vecj^{(n-1)}(E,\veck)$ and the corresponding $\Diff^{(n-1)}(E)$ according to
\begin{align}
\vecj^{(n)}(E,\veck)=\vecj_0(E,&\veck) + \vert \av{\Gr}(E,\veck) \vert^2 \int \frac{\ud \veck'}{(2\pi)^d} \, \vecj^{(n-1)}(E,\veck') \nonumber \\
& \times \left[ U^{\textrm{\tiny B}}_{\veck,\veck'}+{U^{\textrm{\tiny MC}}_{\veck,\veck'}}(\Diff^{(n-1)},E) \right]
\end{align}
and compute $\Diff^{(n)}(E)$ from Eq.~\eqref{eq:kubo}
We first solve the Boltzmann case ($U=U^{\textrm{\tiny B}}$), which we initialize with $\vecj^{(n=0)}=\vecj_0$.
In the full calculation, we then use $\vecj^{(n=0)}=\vecj^{\textrm{\tiny B}}$, which improves the convergence.
The procedure is iterated until convergence of the norm of $\vecj$.

The functions $\vecj(E,\veck)$ are represented in cylindrical coordinates, which are adapted to the symmetry of our problem.
We discretize the integrals on a grid in spherical coordinates, with up to 500 points in the radial direction and 128 points for the angular coordinates.
The different functions appearing in the integral are interpolated using spline interpolation.
We have checked that increasing the grid generates negligible changes in the final results and that our procedure gives consistent results in the isotropic case.

\end{document}